\documentstyle[preprint,aps,tighten,epsf,floats]{revtex}
\newcommand{\nn}{\nonumber\\}
\newcommand{\NNN}{$N\!N\!N$}
\newcommand{\NNNNNN}{$N\!N\!N\rightarrow N\!N\!N$}
\newcommand{\bk}{\bar{k}}
\newcommand{\bq}{\bar{q}}
\newcommand{\bp}{\bar{p}}
\newcommand{\bu}{\bar{u}}
\newcommand{\tPhi}{\tilde{\Phi}}
\newcommand{\tphi}{\tilde{\phi}}
\newcommand{\tX}{\tilde{X}}

\newcommand{\tT}{\tilde{T}}

\newcommand{\ttt}{\tilde{t}}
\newcommand{\td}{\tilde{d}}
\newcommand{\tj}{\tilde{j}}

\newcommand{\la}{\langle}
\newcommand{\ra}{\rangle}
\newcommand{\pslash}{\not\hspace{-0.7mm}p}
\newcommand{\ben}{\begin{displaymath}}
\newcommand{\een}{\end{displaymath}}
\newcommand{\be}{\begin{equation}}
\newcommand{\ee}{\end{equation}}
\newcommand{\bea}{\begin{eqnarray}}
\newcommand{\eea}{\end{eqnarray}}
\newcommand{\eqn}[1]{\label{#1}}
\newcommand{\eq}[1]{Eq.~(\ref{#1})}
\newcommand{\eqs}[1]{Eqs.~(\ref{#1})}
\newcommand{\fign}[1]{\label{#1}}
\newcommand{\fig}[1]{Fig.~\ref{#1}}
\newcommand{\bPsi}{\bar{\Psi}}
\newcommand{\bPhi}{\bar{\Phi}}
\newcommand{\bphi}{\bar{\phi}}
\newcommand{\bfp}{{\bf p}}

\newcommand{\bfq}{{\bf q}}
\newcommand{\bfk}{{\bf k}}
\begin{document}
\draft
\title{Gauging the three-nucleon spectator equation}
\author{A. N. Kvinikhidze\footnote{On leave from Mathematical Institute of
Georgian Academy of Sciences, Tbilisi, Georgia.} and B. Blankleider}
\address{Department of Physics, The Flinders University of South Australia,
Bedford Park, SA 5042, Australia}
\date{\today}
\maketitle

\begin{abstract}
We derive relativistic three-dimensional integral equations describing the
interaction of the three-nucleon system with an external electromagnetic field.
Our equations are unitary, gauge invariant, and they conserve charge. This has
been achieved by applying the recently introduced gauging of equations method to
the three-nucleon spectator equations where spectator nucleons are always on
mass shell. As a result, the external photon is attached to all possible places
in the strong interaction model, so that current and charge conservation are
implemented in the theoretically correct fashion. Explicit expressions are given
for the three-nucleon bound state electromagnetic current, as well as the
transition currents for the scattering processes $\gamma ^3$He$\rightarrow
N\!N\!N$, $Nd\rightarrow \gamma Nd$, and $\gamma ^3$He$\rightarrow Nd$. As a
result, a unified covariant three-dimensional description of the \NNN
-$\gamma$\NNN\ system is achieved.
\end{abstract}

\pacs{21.45.+v, 24.10.Jv, 25.20.-x, 25.30.Bf, 25.30.Fj}

\section{Introduction}

The difficulty of solving four-dimensional scattering equations has led to a
number of three-dimensional reduction schemes that preserve the covariance and
unitarity of the original equations \cite{Log,Sug,Kad,Gross,Erk}. Here we shall
be concerned with one of these schemes, that introduced by Gross \cite{Gross},
where some of the particles, typically the spectator particles of the given
process, are restricted to be on their mass shell. The resultant
three-dimensional equations are called the ``spectator equations''. In the
three-particle system, for example, the spectator particle is well defined (it's
the one flying past two interacting particles), and putting it on mass shell in
every intermediate state results in the three-body spectator equations. The
Gross approach has been used recently in successful relativistic calculations of
nucleon-nucleon scattering \cite{Hol}, elastic electron-deuteron scattering
\cite{Van}, pion photoproduction from the nucleon \cite{Sur}, and the triton
binding energy \cite{Sta}. The quantities used or obtained from these
calculations, such as the three-nucleon bound state wave function, one- and
two-body interaction currents, etc., form just what would be needed to calculate
the electromagnetic properties of the three-nucleon system.  Unfortunately, the
expressions needed to calculate such electromagnetic properties are not
presently available.

The purpose of this paper is therefore to derive, within the framework of the
spectator approach, gauge invariant expressions for the various electromagnetic
transition currents of the three-nucleon system. In particular, we give
expressions for the three-nucleon bound state current from which the triton or
$^3$He electromagnetic form factors follow directly. We also derive expressions
for the scattering processes $\gamma ^3$He$\rightarrow N\!N\!N$, $Nd\rightarrow
\gamma Nd$, and $\gamma ^3$He$\rightarrow Nd$ (here, as in the rest of the
paper, we use $^3$He as the generic symbol for a three-nucleon bound state).

The main tool of the derivation is the method of gauging equations introduced by
us recently for four-dimensional equations \cite{G4d}, and for three-dimensional
equations within the spectator approach \cite{G3d}. This method results in
electromagnetic amplitudes where the external photon is effectively coupled to
every part of every strong interaction diagram in the model. Current and charge
conservation are therefore implemented in the theoretically correct fashion. For
the spectator approach, the gauging of equations method has two especially
important features. Firstly, it avoids the difficulty of choosing the spectator
particles in approaches where the photon is first coupled to hadrons at the
level of four-dimensional quantum field theory. Once the hadronic spectator
equations are specified, the gauging of equations method attaches photons in an
automatic way, without the need for any new spectator particles to be
introduced.  Secondly, when applied to four-dimensional three-nucleon equations,
the gauging of equations method has enabled us to avoid double counting of
diagrams overlooked in previous works \cite{G4d}. This means that in the present
case of the spectator approach, such overcounting is likewise automatically
avoided by the use of the gauging of equations method.

A key ingredient in our final expressions is $\delta^\mu$, the gauged
on-mass-shell propagator for the nucleon. Knowledge of an explicit form for
$\delta^\mu$ that satisfies both the Ward-Takahashi identity and the Ward
identity is essential for the gauge invariance and charge conservation
properties of the three-nucleon electromagnetic currents presented in this
paper. Such a $\delta^\mu$ that satisfies both these identities has been
presented in Refs. \cite{G3d} and \cite{delta^mu}. Thus we have brought together
all the expressions necessary for a covariant, unitary, gauge invariant, and
charge conserving three-dimensional calculation of the electromagnetic
properties of the three-nucleon system.

\section{Gauging the three-nucleon bound state equation}

\subsection{The spectator equation}

In this presentation we work within the framework of the spectator equations for
three identical particles in the absence of three-body forces \cite{Gross}. In
this formalism two of the three particles are restricted to their mass shell by
the following replacement of the usual Feynman propagator $d(p)$
\be
d(p)=\frac{i\Lambda(p)}{p^2-m^2+i\epsilon}\hspace{5mm}
\rightarrow\hspace{5mm}\delta(p)=2\pi \Lambda(p)
\delta^+(p^2-m^2) 
\ee
where $\Lambda(p)=1$ or ${\pslash}+m$ for scalar and spinor particles
respectively, and $\delta^+(p^2-m^2)$ is the positive energy on-mass-shell
$\delta$-function. We refer to $\delta(p)$ as the ``on-mass-shell particle
propagator''.
\begin{figure}[t]
\hspace*{5cm}  \epsfxsize=8cm\epsfbox{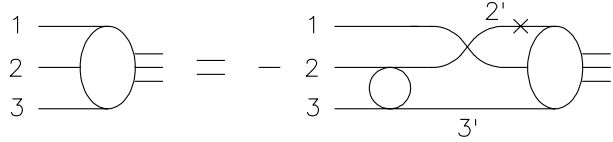}
\vspace{2mm}
\caption{\fign{phi3d} Illustration of \protect\eq{Phi_1} for the bound state
vertex function $\Phi_1$. The on mass shell particle is indicated by a cross.}
\end{figure}

In the four-dimensional formalism of quantum field theory, we may write the
three-body bound state equation in symbolic form as \cite{Sta}
\be
\Phi_1=-t_1DP_{12}\Phi_1   \eqn{Phi_1_bs}
\ee
where $\Phi_1$ is the Faddeev component of the bound-state vertex function
(from now on simply called ``the bound-state vertex function'')
describing the contribution to the bound state from all processes where the (23)
pair interacts last, $t_1$ is the off-shell scattering amplitude of the (23)
pair, $D=d_2d_3$ is the propagator of the (23) pair, and $P_{12}$ is the
operator interchanging particles 1 and 2. Note that our $t_1$ is fully
antisymmetric while $\Phi_1$ is antisymmetric only under the interchange of its
2nd and 3rd particle labels.\footnote{\eq{Phi_1_bs} differs by a factor $-2$
from the corresponding equation in Ref.\ \cite{Sta} due to the use of different
conventions for $t$ and $d$.} Because of these symmetries, one can equally well
use $P_{13}$ in \eq{Phi_1_bs} instead of $P_{12}$ without changing the value of
$\Phi_1$. Once the $P_{12}$ form is chosen as in \eq{Phi_1_bs}, the bound state
``spectator equation'' is obtained from \eq{Phi_1_bs} by putting particle 2 on
the mass shell in intermediate state, i.e. by the replacement
$d_2\rightarrow\delta_2$ in $D$:
\be
\Phi_1=-t_1\delta_2d_3P_{12}\Phi_1   .  \eqn{Phi_1}
\ee
The explicit numerical form of \eq{Phi_1} is given in the Appendix, see
\eq{Phi_1_ex}.  Had we chosen the $P_{13}$ form of the bound state equation, the
spectator equation would instead be defined by putting particle 3 on mass shell;
however, the solution obtained would be identical to that obtained from
\eq{Phi_1}. We illustrate \eq{Phi_1} in \fig{phi3d}. Of course to get a closed
three-dimensional equation for $\Phi_1$ it is necessary to also put the external
particles 1 and 2 on mass shell in \eq{Phi_1}.


It is useful to point out that the spectator equation is not the only
possible three-dimensional equation that follows from \eq{Phi_1_bs} by putting
two particles on mass shell (in three-body intermediate states). However, it is
the best one. Indeed we can investigate all the possibilities by iterating
\eq{Phi_1_bs} once:
\bea
\Phi_1&=&t_1d_2d_3P_{12}t_1d_2d_3P_{12}\Phi_1\nn
&=&t_1d_3P_{12}t_1P_{12}d_1d_2d_3\Phi_1    \eqn{Phi_1_comp}
\eea
thereby obtaining an equation for $\Phi_1$ with the compact kernel
$t_1d_3P_{12}t_1P_{12}=t_1d_3t_2$. \eq{Phi_1_comp} 
shows that there are only three possibilities to restrict two of the 
three intermediate state particles to their mass shells:
\bea
\mbox{(a)}&&\hspace{1cm}\Phi_1=t_1d_3t_2\delta_1\delta_2d_3\Phi_1, \eqn{12}\\
\mbox{(b)}&&\hspace{1cm}\Phi_1=t_1d_3t_2\delta_1d_2\delta_3\Phi_1, \eqn{13}\\
\mbox{(c)}&&\hspace{1cm}\Phi_1=t_1d_3t_2d_1\delta_2\delta_3\Phi_1 .\eqn{23}
\eea
\eq{12} is just the first iteration of the spectator equation \eq{Phi_1}.  After
setting the external particles 1 and 2 on mass shell, the two-body t-matrices in
\eq{12} have two legs on shell and two legs off shell, and therefore depend on
one parameter, the off-mass-shell energy, just like two-body t-matrices in
quantum mechanics.  \eqs{13} and (\ref{23}), on the other hand, are not
iterations of any form similar to \eq{Phi_1} with a kernel linear in
$t_1$. Moreover, after setting two of the external particles on mass shell to
get closed equations, the kernels $t_1d_3t_2$ in \eqs{13} and (\ref{23}) suffer
a major drawback in that one of the t-matrices $t_1$ or $t_2$ has three legs
that are off mass shell. These observation can be seen explicitly in the
illustrations of \eqs{12}-(\ref{23}) given in \fig{3phi_1}.
\begin{figure}[t]
\hspace*{0mm}  \epsfxsize=16.5cm\epsfbox{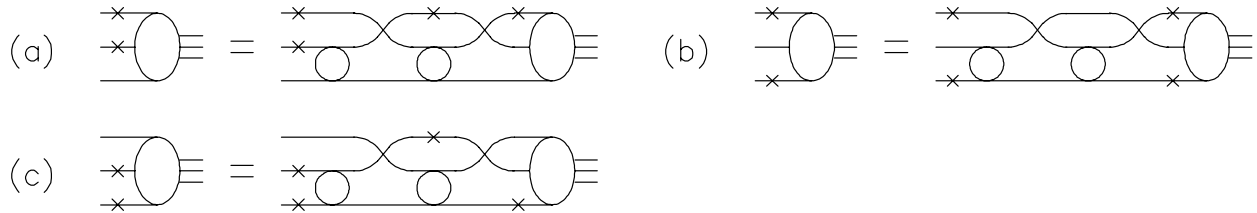}
\vspace{5mm}

\caption{\fign{3phi_1} All the possible three-nucleon bound state equations with
two particles on mass shell. (a)~The spectator equation as given in
\protect\eq{12}. (b) The bound state equation given in \protect\eq{13}.
(c)~The bound state equation given in \protect\eq{23}.}
\end{figure}

\subsection{Gauging the spectator equation}

The question of how to couple an external electromagnetic field to a system of
hadrons described by four-dimensional integral equations, and still retain gauge
invariance, has now been solved. On the two-particle level the problem was first
solved by Gross and Riska \cite{GR} who showed that the one-body current
combined with the gauged interaction kernel of the two-body Bethe-Salpeter
equation gives a gauge invariant two-body current. Similar progress was made by
van Antwerpen and Afnan \cite{AA} who showed how to construct a gauge invariant
current for the relativistic $\pi N$ system where pion absorption can take
place. More recently, we have introduced a general method where any system
described by integral equation can be gauged \cite{G4d}. The method involves the
idea of gauging the integral equations themselves, and results in an
electromagnetic current where the photon is coupled to all possible places in
all possible strong interaction Feynman graphs of the model. We have applied the
gauging of equations method to the relativistic three-nucleon system, thereby
solving an overcounting problem that had previously been overlooked
\cite{G4d}. In this section we would like to apply our method to gauge the bound
state spectator equation, \eq{Phi_1}, in order to obtain a relativistic gauge
invariant three-dimensional description of the three-nucleon bound state
current.

In our procedure, we do not use the on-mass-shell $\delta$-function to eliminate
the zero'th component of the spectator internal momentum in \eq{Phi_1} until
after the gauging of the equation is done. Instead we follow the method outlined
in Ref.~\cite{G3d} and treat \eq{Phi_1} as an eight-dimensional Bethe-Salpeter
equation where some of the propagators are represented by on-mass-shell
$\delta$-functions. This enables us to apply our method of gauging in just the
same way as was done for the eight-dimensional case of \eq{Phi_1_bs} \cite{G4d}.
Gauging \eq{Phi_1} in this way, it immediately follows that
\be
\Phi_1^\mu =-t_1\delta_2d_3P_{12}\Phi_1^\mu-
\left(t_1^\mu\delta_2d_3+t_1\delta^\mu_2d_3+t_1\delta_2d^\mu_3\right)P_{12} 
\Phi_1    . \eqn{Phi_1^mu}
\ee
It is clear from the form of this equation that the quantity $\Phi_1^\mu$
corresponds to that part of the $^3$He $\rightarrow N\!N\!N$ electromagnetic
transition current where the (23) pair was last to interact, and where no
photons are attached to the external constituent legs (a rigorous proof of this
statement was given for the case of four-dimensional quantum field theory in
Ref.~\cite{G4d}). In this respect we note that the bound state vertex component
$\Phi_1$ is a purely nonperturbative object and as such cannot be represented as
a sum of diagrams; nevertheless, $\Phi_1^\mu$ can be formally considered as
$\Phi_1$ with photons attached everywhere ``inside''. Note that \eq{Phi_1^mu} is
an integral equation for $\Phi_1^\mu$ with $\Phi_1$ being an input. Another
input is the gauged Feynman propagator $d_3^{\mu}$. For particle $i=1,2,$ or 3,
the gauged Feynman propagator $d_i^{\mu}$ is defined by
\be
d_i^{\mu}(p',p)=d_i(p')\Gamma_i^{\mu}(p',p)d_i(p)       \eqn{delta^mu}
\ee
where $\Gamma_i^{\mu}(p',p)$ is the particle's electromagnetic vertex
function. For a structureless nucleon of charge $e_i$,
$\Gamma_i^{\mu}(p',p)=e_i\gamma_i^{\mu}$. A further input in
\eq{Phi_1^mu} is $\delta_2^{\mu}$, the gauged on-mass-shell propagator of
particle~2. As shown in Ref.\ \cite{G3d}, taking the explicit form
\be
\delta^{\mu}(p',p)=2\pi i\Lambda(p')\Gamma^{\mu}(p',p)\Lambda(p)\frac
{\delta^+(p'^2-m^2)-\delta^+(p^2-m^2)}{p^2-p'^2}    \eqn{delta_i^mu} 
\ee
for the gauged on-mass-shell propagator $\delta^\mu$, ensures current and charge
conservation of our final results. This is a consequence of the fact that the
$\delta^\mu$ of \eq{delta_i^mu} satisfies both the Ward-Takahashi identity
\be
(p'_{\mu}-p_{\mu})\delta^{\mu}(p',p)=ie\left[\delta(p)-\delta(p')\right]
\eqn{wti}
\ee
as well as the Ward identity
\be
\delta^{\mu}(p,p)=-ie\frac{\partial\delta(p)}{\partial p_\mu}, \eqn{Ward}
\ee
and that \eq{Phi_1^mu} gives an expression for $\Phi_1^\mu$ which has photons
coupled everywhere.

We may formally solve \eq{Phi_1^mu} to obtain
\be
\Phi_1^\mu =-\left(1+t_1\delta_2d_3P_{12}\right)^{-1}
\left(t_1^\mu\delta_2d_3+t_1\delta^\mu_2d_3+t_1\delta_2d^\mu_3\right)P_{12}
\Phi_1 .      \eqn{Phi_1^mu_X}
\ee
The factor $\left(1+t_1\delta_2d_3P_{12}\right)^{-1}$ in this equation 
clearly describes the final state \NNNNNN\ process. Defining
\be
X=\left(1+t_1\delta_2d_3P_{12}\right)^{-1}  \eqn{X_def}
\ee
it follows that $X$ satisfies the two equations
\be
X=1-t_1\delta_2d_3P_{12}X; \hspace{1cm}  X=1-Xt_1\delta_2d_3P_{12}.  \eqn{X}
\ee
As expected, these are three-nucleon scattering equations whose kernel is
identical to that of the bound state equation, \eq{Phi_1}. We illustrate the
first three iterations of these equations in \fig{xfig}. It is evident that
$X$ consists of all possible \NNNNNN\ diagrams where the (13) nucleon pair is
first to interact and the (23) pair is last to interact.
\begin{figure}[t]
\hspace*{1cm}  \epsfxsize=15cm\epsfbox{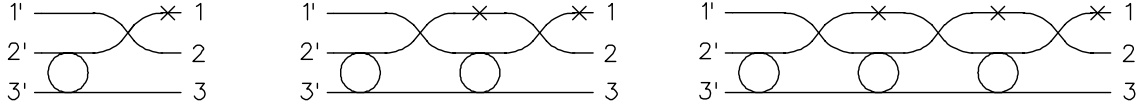}
\vspace{4mm}
\caption{\fign{xfig} Illustration of the first three iterations of the
equations for $X$, \protect\eqs{X}. That it is the spectator particle that is
on mass shell is clearly visible.}
\end{figure}

\subsection{Three-body bound state current}

We recall that $\Phi_1^\mu$ describes the $^3$He $\rightarrow N\!N\!N$
electromagnetic transition current where the (23) nucleon pair is last to
interact, and where no photons are attached to the final state nucleon legs. As
such, it contains all the information that is necessary to specify the
three-nucleon bound state interaction current $j^\mu$. Indeed, we shall use the
expression for $\Phi_1^\mu$ given in \eq{Phi_1^mu_X} to extract $j^\mu$. The key
observation about \eq{Phi_1^mu_X} is that the final state interaction term
$X=(1+t_1\delta_2d_3P_{12})^{-1}$ has a pole at $K^2=M^2$ where $K$ is the total
four-momentum and $M$ is the mass of the three-nucleon bound state. This follows
from the fact that the equations for $X$ and $\Phi_1$ have the same kernel and
that the solution for $\Phi_1$ exists. The three-body bound state current then
follows by taking the residue of \eq{Phi_1^mu_X} at this pole.

We write the pole structure of $X$ as
\be
X(K;p_1p_2,q_1q_2) \sim i \frac{\Phi_1^K(p_1p_2)\bPsi^K_2(q_1q_2)}{K^2-M^2}
\hspace{1cm}\mbox{as}\hspace{1cm}K^2\rightarrow M^2,    \eqn{Xpole}
\ee
which defines the quantity $\bPsi_2$. In order to determine $\bPsi_2$, we take
residues of \eqs{X} at the three-nucleon bound state pole, thereby obtaining
the equations
\be
\Phi_1=-t_1\delta_2d_3P_{12}\Phi_1;\hspace{1cm}
\bPsi_2=-\bPsi_2t_1\delta_2d_3P_{12}.          \eqn{PhiPsi}
\ee 
The first of these is the bound state equation for $\Phi_1$, which of course is
the reason that Eq.(\ref{Xpole}) was written with a $\Phi_1$ factor. The second
equation can be written as
\be
\bPsi_2=-\bPsi_2P_{12}t_2\delta_1d_3     \eqn{bPsi_2}
\ee
which has the same form as the equation for the second Faddeev component of the
bound state wave function in four-dimensional quantum field theory \cite{G4d},
hence our choice of notation for $\bPsi_2$. However, in contrast to the
four-dimensional quantum field theory case, the $\bPsi_2$ of \eq{bPsi_2}
contains explicit on-mass shell propagators.  This can already be seen from
\eq{bPsi_2} where the $\delta_1$ that is present on the r.h.s. contains a
$\delta$-function that is not integrated over. But the full structure of
$\bPsi_2$ becomes clear only after we iterate \eq{bPsi_2} once, obtaining
\be
\bPsi_2=\bPsi_2 P_{12}t_2\delta_1d_3 P_{12}t_2\delta_1d_3 
= \bPsi_2 P_{12}t_2d_3 t_1P_{12}\delta_1\delta_2d_3 .\eqn{iterate}
\ee
This reveals an explicit factor $\delta_1\delta_2d_3$ with two on-mass-shell
propagators, followed by the connected term $t_2d_3 t_1$. Thus $\bPsi_2$ has a
structure of the form
\be
\bPsi_2 = -\bPhi_1P_{12}\delta_1\delta_2d_3   \eqn{bPsi_2_structure}
\ee
where
\be
\bPhi_1 = -\bPsi_2 P_{12}t_2d_3 t_1
\ee
has no propagators on its three external legs. Multiplying \eq{bPsi_2} on the
right by $-P_{12}t_2d_3 t_1$, we find that $\bPhi_1$ satisfies the equation
\be
\bPhi_1=-\bPhi_1 P_{12}\delta_2d_3 t_1.    \eqn{bPhi_1}
\ee
This is the conjugate equation to \eq{Phi_1}, hence our choice of notation for
$\bPhi_1$ in \eq{bPsi_2_structure}.  With $\bPhi_1$ and $\bPsi_2$ determined by
\eqs{bPsi_2_structure} and (\ref{bPhi_1}), the residue of \eq{Xpole} is
completely specified.

Thus, in the vicinity of the three-body bound state pole ($K^2\rightarrow M^2$),
$\Phi_1^\mu$ behaves as
\be
\Phi_1^\mu \sim
-i\frac{\Phi_1\bPhi_1P_{12}\delta_1\delta_2d_3 }{K^2-M^2}
\left(t_1^\mu\delta_2d_3+t_1\delta^\mu_2d_3+t_1\delta_2d^\mu_3\right)
P_{12}\Phi_1 .      \eqn{Phi_1^mu_pole}
\ee
The three-nucleon bound state current in quantum field theory is given by the
matrix element $\la K|J^\mu(0)|Q\ra$ of the electromagnetic current operator
$J^\mu$ between momentum eigenstates $|K\ra$ and $|Q\ra$. In the spectator
approximation it can be determined by taking the residue of \eq{Phi_1^mu_pole}
at the three-nucleon bound state pole on the left:
\be
\la K|J^\mu(0)|Q\ra \equiv j^\mu(K,Q) = \bPhi_{1}^K P_{12}\delta_1\delta_2d_3
\left(t_1^\mu\delta_2d_3+t_1\delta^\mu_2d_3+t_1\delta_2d^\mu_3\right)
P_{12}\Phi_1^Q .    \eqn{jjj}
\ee
Here $K$ and $Q$ are the total four-momenta of the final and initial bound
states, respectively, with $K^2=Q^2=M^2$ and $K=Q+q$ where $q$ is the
four-momentum of the incoming photon. One can eliminate $t_1$ from this
expression by using \eq{bPhi_1}, in this way obtaining
\be
j^\mu(K,Q) = \bPhi_1^K P_{12}\delta_1\delta_2d_3
t_1^\mu\delta_2d_3P_{12}\Phi_1^Q-\bPhi_1^K\delta_1\left(\delta^\mu_2d_3+
\delta_2d^\mu_3\right)P_{12}\Phi_1^Q. \eqn{j^mu}
\ee
This expression is illustrated in \fig{jmu3d}. Note that the last two terms
do not give the full one-body contribution to the bound state current as
a further contribution comes from the gauged propagators inside $t_1^\mu$.
\begin{figure}[t]
\hspace*{1cm} \epsfxsize=15cm\epsfbox{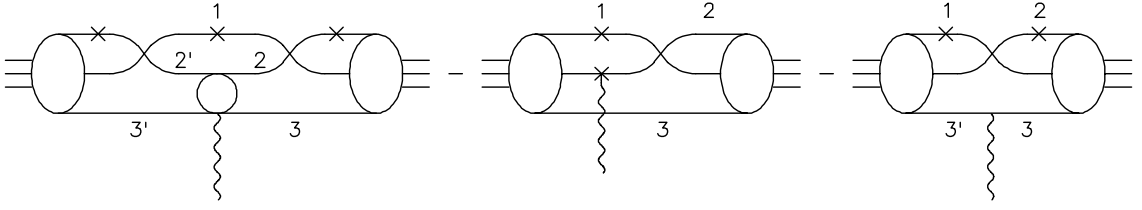}
\vspace{2mm}
\caption{\fign{jmu3d} The three-nucleon bound state current in the spectator
model as given by \protect\eq{j^mu}. Particle labels correspond to those used in
\protect\eq{j^mu-explicit}.}
\end{figure}

To find $t_1^\mu$, we first need to specify the spectator equations for $t_1$:
\be
t_1=v_1+\frac{1}{2}v_1\delta_2d_3t_1;
\hspace{1cm}t_1=v_1+\frac{1}{2}t_1\delta_2d_3v_1.     \eqn{t_1}
\ee
By gauging these equations
one can express $t_1^\mu$ in terms of the interaction current $v_1^\mu$ as
\be
t_1^\mu=\frac{1}{2}t_1\left(\delta^\mu_2d_3+\delta_2d^\mu_3\right)t_1+
\left(1+\frac{1}{2}t_1\delta_2d_3\right)v_1^\mu
\left(1+\frac{1}{2}\delta_2d_3t_1\right). \eqn{t_1^mu}
\ee
Note that our $v_1$ is the sum of all possible irreducible diagrams for the
scattering of two identical particles, therefore $P_{23}v_1=v_1P_{23}=-v_1$.
That is why we do not need to use the symmetrised propagator
$\frac{1}{2}(\delta_2d_3+d_2\delta_3)$ in \eq{t_1} in order to satisfy
the  Pauli exclusion principle.

Although \eq{j^mu} may be the most practical equation for numerical
calculations, with the help of \eq{t_1^mu} we can also eliminate $t_1^\mu$
in favour of the interaction current $v_1^\mu$:
\be
j^\mu(K,Q) = \bPhi_1^K\delta_1\left(\delta^\mu_2d_3+\delta_2d^\mu_3\right)
\left(\frac{1}{2}-P_{12}\right)\Phi_1^Q +
\bPhi_1^K\left(P_{12}-\frac{1}{2}\right)\delta_2d_3\delta_1v_1^\mu \delta_2d_3
(P_{12}-\frac{1}{2})\Phi_1^Q . \eqn{j^mu-v}
\ee
It is interesting to compare \eq{j^mu-v} with the corresponding expression
obtained by using the same gauging method in the case of four-dimensional
quantum field theory \cite{G4d}:
\be
j^\mu(K,Q) = \bPhi_1^Kd_1\left(d^\mu_2d_3+d_2d^\mu_3\right)
\left(\frac{1}{2}-P_{12}\right)\Phi_1^Q
+\bPhi_1^K\left(P_{12}-\frac{1}{2}\right)d_2d_3d_1v_1^\mu
d_2d_3(P_{12}-\frac{1}{2}) \Phi_1^Q . \eqn{j^mu-v4d}
\ee
This comparison makes clear the prescription $d_1\rightarrow\delta_1$,
$d_2\rightarrow\delta_2$, $d^\mu_1\rightarrow\delta^\mu_1$,
$d^\mu_2\rightarrow\delta^\mu_2$ that one should use to obtain the three-body
bound state electromagnetic current in the three-dimensional spectator approach,
\eq{j^mu-v}, from the corresponding four-dimensional expression of
Eq.~(\ref{j^mu-v4d}).

In the impulse approximation where the interaction current $v_1^\mu$ is
neglected, we have that
\be
j^\mu(K,Q) =
\bPhi_1^K\delta_1\left(\delta^\mu_2d_3+\delta_2d^\mu_3\right)
\left(\frac{1}{2}-P_{12}\right)\Phi_1^Q . \eqn{j^mu-impulse}
\ee
This of course is the full one-body contribution to the bound state current.
Because of propagator $\delta_1$ in this expression, particle 1 is on mass shell
(of course to the right of operator $P_{12}$ this on-mass shell particle becomes
particle 2). The first term on the r.h.s of \eq{j^mu-impulse} also contains the
gauged propagator $\delta^\mu_2$, and therefore according to \eq{delta_i^mu},
particle 2 can be off mass shell either to the left or to the right of the
photon. Thus to calculate this first term, one needs to know $\bPhi_1^K$ and
$\Phi_1^Q$ where only one external particle is on mass shell.  These can always
be determined from the spectator bound state vertex functions where two
particles are on mass shell by using \eq{Phi_1} and \eq{bPhi_1} Choosing the
momenta of particles 1 and 2 as independent variables, we may write
\eq{j^mu-impulse} in the explicit numerical form

\bea
j^\mu(K,Q) &=&\int\frac{d^4p_1}{(2\pi^4)}\frac{d^4p_2}{(2\pi)^4}
\bar{\Phi}_1^K(\bp_1,p_2+q)\delta(p_1)
\delta^\mu(p_2+q,p_2)
d(Q-p_1-p_2)
\left(\frac{1}{2}-P_{12}\right)\Phi_1^Q(\bp_1,p_2)\nn
+&\displaystyle\int& \frac{d^4p_1}{(2\pi)^4}\frac{d^4p_2}{(2\pi)^4}
\bar{\Phi}_1^K(\bp_1,\bp_2)\delta(p_1)\delta(p_2)
d^\mu(K-p_1-p_2,Q-p_1-p_2)\left(\frac{1}{2}-P_{12}\right)
\Phi_1^Q(\bp_1,\bp_2)  \nn \eqn{26}
\eea
where the momenta which are on-mass-shell are labelled with a bar over the top.

\subsection{Gauge invariance}

As the gauging of equations method effectively couples photons everywhere in
the strong interaction model, gauge invariance is guaranteed.  Nevertheless,
here we would like to check this explicitly on our derived expression for the
bound state current of \eq{j^mu}.

Writing this equation out in full numerical form we have that
\bea
j^\mu(K,Q) &=& \int \frac{d^4p_1}{(2\pi)^4}\frac{d^4p_{2}}{(2\pi)^4}
\frac{d^4p'_{2}}{(2\pi)^4}\,
\bPhi_1^K(p'_2p_1p'_3) \delta(p_1)\delta(p'_2)d(p'_3)
t^\mu(p'_2p'_3,p_2p_3)\delta(p_2)d(p_3)\Phi_1^Q(p_2p_1p_3)\nn
&-&\int \frac{d^4p_1}{(2\pi)^4}\frac{d^4p_{2}}{(2\pi)^4}\,
\bPhi_1^K(p_1p'_2p_3)\delta(p_1)
\delta^\mu(p'_2,p_2)d(p_3)\Phi_1^Q(p_2p_1p_3)\nn
&-&\int \frac{d^4p_1}{(2\pi)^4}\frac{d^4p_{2}}{(2\pi)^4}\,
\bPhi_1^K(p_1p_2p'_3)\delta(p_1)
\delta(p_2)d^\mu(p'_3,p_3)\Phi_1^Q(p_2p_1p_3) , \eqn{j^mu-explicit}
\eea
where $p_3=Q-p_1-p_2$, $p'_3=K-p_1-p_2'$, and it is understood that
$p'_2+p'_3=p_2+p_3+q$ in the first integral, $p'_2=p_2+q$ in the second
integral, and $p'_3=p_3+q$ in the third.  Here we have followed the notation of
Ref.~\cite{G4d} and displayed the momentum of each particle explicitly. Each of
the gauged inputs in \eq{j^mu-explicit} satisfies a Ward-Takahashi identity
(WTI). In the notation of \eq{j^mu-explicit}, the WTI for $t^\mu$ takes the form
\bea
q_\mu t^\mu(p'_2p'_3,p_2p_3) &=& i
\left[e_2t(p'_2-q,p'_3;p_2p_3)-t(p'_2p'_3;p_2+q,p_3)e_2\right]\nn
&+& i
\left[ e_3t(p'_2,p'_3-q;p_2p_3)-t(p'_2p'_3;p_2,p_3+q)e_3\right] ,
\eqn{ward-t}
\eea
while for $\delta^\mu$ and $d^\mu$ the WTI's are
\bea
q_\mu \delta^{\mu}(p'_2,p_2)&=&ie_2\left[\delta(p_2)-\delta(p'_2)\right],\\
\eqn{ward-delta}
q_\mu d^{\mu}(p'_3,p_3)&=&ie_3\left[d(p_3)-d(p'_3)\right]. \eqn{ward-d}
\eea
In the present case of three nucleons, the charges $e_i$ ($i=1,2,3$)
are given by
$e_i=\frac{1}{2}[1+\tau_3^{(i)}]e_p$ where $\tau_3$ is the Pauli matrix for the
third component of isospin, and $e_p$ is the charge of the proton.

In order to prove gauge invariance of the bound state current, we follow the
same procedure as we used for the distinguishable particle case \cite{G4d}, and
evaluate the quantity $q_\mu j^{\mu}$ by using the above WTI's in
\eq{j^mu-explicit}. However, unlike in the distinguishable particle case, subtle
use of identical particle symmetry also needs to be made before the final
expression is reduced to zero. Although this is straightforward, working with
lengthy numerical expression like that of \eq{j^mu-explicit} tends to obscure
the presentation.  For this reason, here we would prefer to avoid the use of
explicit numerical forms and instead to keep all our equations at the symbolic
level. In order to write the WTI's of Eqs.~(\ref{ward-t})-(\ref{ward-d}) in
symbolic form, we introduce the quantities $\hat{e}_i$ whose numerical form
is defined by
\be
\hat{e}_i(p'_1p'_2p'_3,p_1p_2p_3) =
ie_i(2\pi)^{12}\delta^4(p'_i-p_i-q)\delta^4(p'_j-p_j)\delta^4(p'_k-p_k),
\ee
where $ijk$ are cyclic permutations of $123$.
Then the above WTI's can be written symbolically in terms of commutators as
\be
q_\mu t_1^\mu=[\hat{e}_2,t_1]+[\hat{e}_3,t_1],\hspace{1cm}
q_\mu \delta_2^\mu=[\hat{e}_2,\delta_2],\hspace{1cm}
q_\mu d_3^\mu=[\hat{e}_3,d_3].
\ee
Using these, the divergence of the three-nucleon bound state current is given by
\bea
\lefteqn{q_\mu j^{\mu} =
\bPhi_1^KP_{12}\delta_1\delta_2d_3
\left([\hat{e}_2,t_1]+[\hat{e}_3,t_1]\right)\delta_2d_3P_{12} \Phi_1^Q}
\hspace{3cm}\nn
&&-\bPhi_1^K\delta_1
\left([\hat{e}_2,\delta_2]d_3+\delta_2[\hat{e}_3,d_3]\right)P_{12}\Phi_1^Q.
\eqn{q.J}  
\eea
Using the bound state equations, \eqs{Phi_1} and (\ref{bPhi_1}), and the fact
that $[\hat{e}_3,\delta_2]=[\hat{e}_3,P_{12}]=0$, \eq{q.J} reduces immediately
down to
\be
q_\mu j^{\mu} =
-\bPhi_1^KP_{12}\delta_1\delta_2d_3\hat{e}_2 \Phi_1^Q
+\bPhi_1^K\delta_1 \delta_2\hat{e}_2d_3P_{12}\Phi_1^Q.   \eqn{q.J_obscure}
\ee
Since $[\hat{e}_2,P_{12}]\ne 0$, it is not immediately obvious that the
last two terms cancel. To show that this is indeed the case, we
make use of the fact that
\be
\hspace{1cm}[\hat{e}_2,P_{12}t_1P_{12}]=[\hat{e}_2,t_2]=0.
\ee
Then using the bound state equation in the last term of \eq{q.J_obscure}
we obtain that
\bea
\bPhi_1^K\delta_1 \delta_2\hat{e}_2d_3P_{12}\Phi_1^Q
&=&-\bPhi_1^K\delta_1 \delta_2\hat{e}_2d_3P_{12}t_1\delta_2d_3P_{12}\Phi_1^Q
= -\bPhi_1^K\delta_1 d_3\delta_2\hat{e}_2P_{12}t_1P_{12}\delta_1d_3\Phi_1^Q\nn
&=& -\bPhi_1^K\delta_1 d_3P_{12}t_1P_{12}\delta_2\hat{e}_2\delta_1d_3\Phi_1^Q
= -\bPhi_1^KP_{12}\delta_2 d_3t_1P_{12}\delta_2\hat{e}_2\delta_1d_3\Phi_1^Q\nn
&=& \bPhi_1^KP_{12}\delta_2\hat{e}_2\delta_1d_3\Phi_1^Q
= \bPhi_1^KP_{12}\delta_1\delta_2d_3\hat{e}_2\Phi_1^Q.
\eea
Using this result in \eq{q.J_obscure} we obtain the current conservation
relation
\be
q_\mu j^{\mu} = 0.
\ee

\subsection{Normalization condition} 

The method for obtaining the normalization condition for bound state wave
functions in quantum field theory typically involves the taking of residues of
Green functions or t-matrices at the bound state pole, and is similar to what is
used in quantum mechanics when the potentials are energy dependent. Here we
apply the same idea, but to the quantity $X$, in order to determine the specific
normalisation condition for the three-body bound state vertex function in the
spectator approach.  Our starting point is the following identity for $X$:
\be
X\left(1+t_1\delta_2d_3P_{12}\right)X=X.    \eqn{XX}
\ee
Using the pole behaviour of $X$ given by \eq{Xpole}, we see that 
in the vicinity of the three-body bound state pole, \eq{XX} reduces to
\be
i \frac{\bPsi_2^Q(1+t_1\delta_2d_3P_{12})\Phi_1^Q}{Q^2-M^2}=1,
\ee
or
\be
\left. i \bPsi_2^Q\frac{\partial(1+t_1\delta_2d_3P_{12})}{\partial Q^2}
\Phi_1^Q\right|_{Q^2=M^2}=1.    \eqn{norm1}
\ee
With the understanding that $Q^2=M^2$, and using \eq{bPsi_2_structure},
this may also be written as 
\be
-i\bPhi_1^QP_{12}\delta_1\delta_2d_3\frac{\partial(t_1\delta_2d_3P_{12})}
{\partial Q^2}\Phi_1^Q=1,
\ee
or
\be
-i\bPhi_1^QP_{12}\delta_1\delta_2d_3\frac{\partial t_1}
{\partial Q^2}\delta_2d_3P_{12}\Phi_1^Q
+i\bPhi_1^Q\delta_1\frac{\partial(\delta_2d_3)}{\partial Q^2}P_{12}\Phi_1^Q=1.
       \eqn{norm-symbolic}
\ee
This form of the normalisation condition is especially convenient as it
is expressed in terms of the two-body t-matrix $t_1$ rather than the potential
$v_1$ which results, for example, when the full Green function is used in an
identity, similar to \eq{XX}, but involving two-body potentials
 \cite{G4d}.

It is sometimes convenient to express the normalization condition as a
four-vector relation by using the replacement
\ben
\frac{\partial}{\partial Q^\mu}\rightarrow 2Q_\mu\frac{\partial}{\partial Q^2}
\een
in the above equations. That this replacement is valid can be easily justified
by appealing to Lorentz invariance.
In this way \eq{norm-symbolic}, written out in full
numerical form, becomes
\bea
&&\int \frac{d^4k_1}{(2\pi)^4}\frac{d^4k_2}{(2\pi)^4}
\frac{d^4p_2}{(2\pi)^4}\,
\bPhi_1^QP_{12}(k_1,k_2)\delta(k_1)\delta(k_2)d(k_3)
\frac{\partial t(Q-k_1,k_2,p_2)}{\partial Q_\mu}
\delta(p_2)d(p_3)P_{12}\Phi_1^Q(k_1,p_2)\nn[3mm]
&-&\int \frac{d^4k_1}{(2\pi)^4}\frac{d^4k_2}{(2\pi)^4}\,\bPhi_1^Q(k_1,k_2)
\delta(k_1)\delta(k_2)\frac{\partial d(k_3)}{\partial Q_\mu}
P_{12}\Phi_1^Q(k_1,k_2) = 2iQ^\mu    \eqn{norm}
\eea
where $k_3=Q-k_1-k_2$ and $p_3=Q-k_1-p_2$.

\subsection{Charge conservation} 

In its usual meaning, charge conservation is a consequence of current
conservation. As we have proved current conservation above, charge is naturally
conserved in our model. On the other hand, that the conserved charge is equal to
the total charge of the physical system does not follow automatically from
current conservation, and therefore needs to be checked separately. In
particular, what needs to be checked is that
\be
j^\mu(Q,Q) = 2e Q^\mu        \eqn{charge}
\ee
where e is the physical charge of the three-body bound state. We follow current
terminology and also refer to Eq.~(\ref{charge}) as a statement of ``charge
conservation'' (in the sense that if $e$ is indeed the physical charge, then no
charge has been ``lost'' in the model).  For an exact solution of field theory,
\eq{charge} follows from the fact that $|Q\ra$ in \eq{jjj} is an eigenstate of
the charge operator with eigenvalue $e$. In this subsection we show that
\eq{charge} also holds in our model where the gauging of equations method has
been used for the spectator approach.

The bound state current was given in its explicit form in \eq{j^mu-explicit}.
We can rewrite this expression for zero momentum transfer, and using only
independent momentum variables, as
\bea
j^{\mu}(Q,Q) &=&\int \frac{d^4k_1}{(2\pi)^4}\frac{d^4k_2}{(2\pi)^4}
\frac{d^4p_2}{(2\pi)^4} 
\bPhi_1^QP_{12}(k_1,k_2)\delta(k_1)\delta(k_2)d(k_3)
t^\mu(Q-k_1,Q-k_1;k_2,p_2)\nn[-2mm]
&&\hspace{5cm}\delta(p_2)d(p_3)P_{12}\Phi_1^Q(k_1,p_2)\nn[2mm]
&-&\int \frac{d^4k_1}{(2\pi)^4}\frac{d^4k_2}{(2\pi)^4}
\bPhi_1^Q(k_1,k_2)\delta(k_1)\delta^\mu(k_2,k_2)
d(k_3)P_{12}\Phi_1^Q(k_1,k_2)\nn
&-&\int \frac{d^4k_1}{(2\pi)^4}\frac{d^4k_2}{(2\pi)^4}
\bPhi_1^Q(k_1,k_2)\delta(k_1)\delta(k_2)d^\mu(k_3,k_3)
P_{12}\Phi_1^Q(k_1,k_2)       \eqn{j^mu-charge}
\eea
where \hspace{1mm} $k_3=Q-k_1-k_2$,\hspace{2mm} $p_3=Q-k_1-p_2$, \hspace{1mm}
and where $t^\mu$ is expressed in terms of the
total momenta in the (23) system, $Q-k_1$ in both the initial and final states,
and the momenta of particle 2, $p_2$ and $k_2$ for initial and final states, 
respectively.

Both the gauged Feynman propagator $d^\mu$ and the gauged on-mass-shell
particle propagator $\delta^\mu$ satisfy the Ward identity \cite{delta^mu}:
\be
id^{\mu}(k_3,k_3)=e_3\frac{\partial d(k_3)}{\partial k_{3\mu}} \eqn{ward_d},
\ee
\be
i\delta^{\mu}(k_2,k_2)=e_2\frac{\partial \delta(k_2)}{\partial k_{2\mu}}
\eqn{ward_delta}.
\ee
The interaction current $v^\mu$ is an input to our model and therefore satisfies
the two-particle Ward identity by construction. In turn, it can easily be shown
that the $t^\mu$, as given by \eq{t_1^mu}, must also satisfy the two-particle
Ward identity. For the momentum variables of \eq{j^mu-charge}, this identity
reads
\bea
it^{\mu}(Q-k_1,Q-k_1;k_2,p_2)&=&
e_2\frac{\partial t(Q-k_1,k_2,p_2)}{\partial k_{2\mu}}+
\frac{\partial t(Q-k_1,k_2,p_2)}{\partial p_{2\mu}}e_2\nn
&+&(e_3+e_2)\frac{\partial t(Q-k_1,k_2,p_2)}{\partial Q_{\mu}}. \eqn{ward_t}
\eea
Substituting \eq{ward_t} into \eq{j^mu-charge}, we may then use the bound state
equations for $\Phi_1^Q$ and $\bPhi_1^Q$ to simplify the terms containing
$\partial t/\partial k_{2\mu}$ and $\partial t/\partial p_{2\mu}$. Writing
\eq{ward_d} as
\be
id^\mu(k_3,k_3)=(e_3+e_2)\frac{\partial d(k_3)}{\partial k_{3\mu}}
-e_2\frac{\partial d(k_3)}{\partial k_{3\mu}}=
(e_3+e_2)\frac{\partial d(k_3)}{\partial Q_\mu}+
e_2\frac{\partial d(k_3)}{\partial k_{2\mu}}
\ee
we may then use it together with \eq{ward_delta} in \eq{j^mu-charge} to obtain
\bea
ij^{\mu}(Q,Q) &=&\int \frac{d^4k_1}{(2\pi)^4}\frac{d^4k_2}{(2\pi)^4}
\frac{d^4p_2}{(2\pi)^4} 
\,\bPhi_1^QP_{12}(k_1,k_2)\delta(k_1)\delta(k_2)d(k_3)(e_3+e_2)
\frac{\partial t(Q-k_1,k_2,p_2)}{\partial Q_{\mu}}\nn[-2mm]
&&\hspace{5cm}\delta(p_2)d(p_3)P_{12}\Phi_1^Q(k_1,p_2)\nn[2mm]
&-&\int \frac{d^4k_1}{(2\pi)^4}\frac{d^4k_2}{(2\pi)^4}\,\bPhi_1^Q(k_1,k_2)
\delta(k_1)\delta(k_2)(e_3+e_2)
\frac{\partial d(k_3)}{\partial Q_\mu}
P_{12}\Phi_1^Q(k_1,k_2)\nn
&-&\int \frac{d^4k_1}{(2\pi)^4}\frac{d^4k_2}{(2\pi)^4}\,\bPhi_1^QP_{12}(k_1,k_2)
\delta(k_1)\delta(k_2)d(k_3)e_2
\frac{\partial \Phi_1^Q(k_1,k_2)}{\partial k_{2\mu}}\nn
&-&\int \frac{d^4k_1}{(2\pi)^4}\frac{d^4k_2}{(2\pi)^4}
\,\frac{\partial\bPhi_1^Q(k_1,k_2)}{\partial k_{2\mu}}e_2
\delta(k_1)\delta(k_2)d(k_3)P_{12}\Phi_1^Q(k_1,k_2)\nn
&-&\int \frac{d^4k_1}{(2\pi)^4}\frac{d^4k_2}{(2\pi)^4}
\,\bPhi_1^Q(k_1,k_2)\delta(k_1)e_2
\frac{\partial \delta(k_2)}{\partial k_{2\mu}}d(k_3)P_{12}\Phi_1^Q(k_1,k_2)\nn
&-&\int \frac{d^4k_1}{(2\pi)^4}\frac{d^4k_2}{(2\pi)^4}
\,\bPhi_1^Q(k_1,k_2)\delta(k_1)\delta(k_2)
e_2\frac{\partial d(k_3)}{\partial k_{2\mu}}P_{12}\Phi_1^Q(k_1,k_2).
\eea
Using integration by parts, we can write the last three terms of this equation
as
\ben 
\int \frac{d^4k_1}{(2\pi)^4}\frac{d^4k_2}{(2\pi)^4}\,\bPhi_1^Q(k_1,k_2)
\delta(k_1)\delta(k_2)
e_2d(k_3)\frac{\partial P_{12}\Phi_1^Q(k_1,k_2)}{\partial k_{2\mu}}.
\een
\eq{j^mu-charge} can then be written as
\bea
ij^{\mu}(Q,Q) &=&\int \frac{d^4k_1}{(2\pi)^4}\frac{d^4k_2}{(2\pi)^4}
\frac{d^4p_2}{(2\pi)^4}\,
\bPhi_1^QP_{12}(k_1,k_2)\delta(k_1)\delta(k_2)d(k_3)(e_3+e_2+e_1)
\frac{\partial t(Q-k_1,k_2,p_2)}{\partial Q_{\mu}}\nn[-2mm]
&&\hspace{6cm}\delta(p_2)d(p_3)P_{12}\Phi_1^Q(k_1,p_2)\nn[3mm]
&-&\int \frac{d^4k_1}{(2\pi)^4}\frac{d^4k_2}{(2\pi)^4}\,\bPhi_1^Q(k_1,k_2)
\delta(k_1)\delta(k_2)(e_3+e_2+e_1)\frac{\partial d(k_3)}{\partial Q_\mu}
P_{12}\Phi_1^Q(k_1,k_2)\nn
&+&
\int \frac{d^4k_1}{(2\pi)^4}\frac{d^4k_2}{(2\pi)^4}\frac{d^4p_2}{(2\pi)^4} \,
\bPhi_1^QP_{12}(k_1,k_2)\delta(k_1)\delta(k_2)d(k_3)e_1
\frac{\partial t(Q-k_1,k_2,p_2)}{\partial k_{1\mu}}\nn[-2mm]
&&\hspace{6cm}\delta(p_2)d(p_3)P_{12}\Phi_1^Q(k_1,p_2)\nn[3mm]
&-&\int \frac{d^4k_1}{(2\pi)^4}\frac{d^4k_2}{(2\pi)^4}\,\bPhi_1^Q(k_1,k_2)
\delta(k_1)\delta(k_2)e_1\frac{\partial d(k_3)}{\partial k_{1\mu}}
P_{12}\Phi_1^Q(k_1,k_2)\nn
&-&\int \frac{d^4k_1}{(2\pi)^4}\frac{d^4k_2}{(2\pi)^4}\,
\bPhi_1^QP_{12}(k_1,k_2)\delta(k_1)\delta(k_2)d(k_3)e_2
\frac{\partial \Phi_1^Q(k_1,k_2)}{\partial k_{2\mu}}\nn
&+&\int \frac{d^4k_1}{(2\pi)^4}\frac{d^4k_2}{(2\pi)^4}\,
\bPhi_1^Q(k_1,k_2)\delta(k_1)\delta(k_2)
d(k_3)e_2\frac{\partial P_{12}\Phi_1^Q(k_1,k_2)}{\partial k_{2\mu}} \eqn{kill4}
\eea
where the charge in the first two terms has been increased to the total charge
of the system, and where we used the fact that \mbox{$\partial
t(Q-k_1,k_2,p_2)/\partial Q_\mu=- \partial t(Q-k_1,k_2,p_2)/\partial k_{1\mu}$}
and \mbox{$\partial d(k_3)/\partial Q_\mu= -\partial d(k_3)/\partial
k_{1\mu}$}. Since the bound state vertex function $\Phi_1^Q$ is an eigenstate of
the total charge $e_1+e_2+e_3$ with eigenvalue $e$, a comparison with
the normalisation condition, \eq{norm}, shows that the first two terms of the
above equation give the sought after charge conservation relation.  Thus all we
need to show now is that the last four terms of \eq{kill4} cancel each other. To
this end we eliminate $t$ in the third term on the r.h.s. of \eq{kill4} by using
integration by parts, and then making use of the bound state equations for
$\Phi_1^Q$ and $\bPhi_1^Q$. In this way we get
\bea
\int &&\frac{d^4k_1}{(2\pi)^4}\frac{d^4k_2}{(2\pi)^4}\frac{d^4p_2}{(2\pi)^4}\,
\bPhi_1^QP_{12}(k_1,k_2)\delta(k_1)\delta(k_2)d(k_3)e_1
\frac{\partial t(Q-k_1,k_2,p_2)}{\partial k_{1\mu}}
\delta(p_2)d(p_3)P_{12}\Phi_1^Q(k_1,p_2)\nn
&&=
\int \frac{d^4k_1}{(2\pi)^4}\frac{d^4k_2}{(2\pi)^4} \frac{\partial\,
\bPhi_1^QP_{12}(k_1,k_2)\delta(k_1)\delta(k_2)d(k_3)}
{\partial k_{1\mu}}e_1\Phi_Q(k_1,k_2)\nn
&&+\int \frac{d^4k_1}{(2\pi)^4}\frac{d^4p_2}{(2\pi)^4}\,
\bPhi_1^Q(k_1,p_2)\delta(k_1)\delta(p_2)e_1
\frac{\partial d(p_3)P_{12}\Phi_Q(k_1,p_2)}{\partial k_{1\mu}}  \eqn{two-terms}
\eea
where $p_2$ and $p_3$ in the last equation can now be replaced by $k_2$ and
$k_3$, respectively. That the last two terms of \eq{two-terms} cancel
the last three terms of \eq{kill4} can then be seen by using the identities
\bea
 &&\int\frac{d^4k_1}{(2\pi)^4}\frac{d^4k_2}{(2\pi)^4} \,
\bPhi_1^Q(k_1,k_2)\delta(k_1)\delta(k_2)e_1
\frac{\partial d(k_3)P_{12}\Phi_1^Q(k_1,k_2)}{\partial k_{1\mu}}\nn
-&&\int \frac{d^4k_1}{(2\pi)^4}\frac{d^4k_2}{(2\pi)^4}\,
\bPhi_1^Q(k_1,k_2)\delta(k_1)\delta(k_2)e_1
\frac{\partial d(k_3)}{\partial k_{1\mu}}
P_{12}\Phi_1^Q(k_1,k_2)\nn
-&&\int \frac{d^4k_1}{(2\pi)^4}\frac{d^4k_2}{(2\pi)^4}\,
\bPhi_1^QP_{12}(k_1,k_2)\delta(k_1)\delta(k_2)d(k_3)e_2
\frac{\partial \Phi_1^Q(k_1,k_2)}{\partial k_{2\mu}} = 0.
\eea
and
\bea
&&\int \frac{d^4k_1}{(2\pi)^4}\frac{d^4k_2}{(2\pi)^4}\,
\frac{\partial\,\bPhi_1^QP_{12}(k_1,k_2)\delta(k_1)\delta(k_2)d(k_3)}
{\partial k_{1\mu}}e_1\Phi_1^Q(k_1,k_2)\nn
+&&\int \frac{d^4k_1}{(2\pi)^4}\frac{d^4k_2}{(2\pi)^4}\,
\bPhi_1^Q(k_1,k_2)\delta(k_1)\delta(k_2)
d(k_3)e_2\frac{\partial P_{12}\Phi_1^Q(k_1,k_2)}{\partial k_{2\mu}}\nn
=&&
\int \frac{d^4k_1}{(2\pi)^4}\frac{d^4k_2}{(2\pi)^4}\,
\frac{\partial\,\bPhi_1^QP_{12}(k_1,k_2)\delta(k_1)\delta(k_2)
d(k_3)e_1\Phi_1^Q(k_1,k_2)}{\partial k_{1\mu}} = 0.
\eea
Thus we have shown \eq{charge}, which proves charge conservation for our gauged
three-nucleon spectator model.

\section{Gauging the three-nucleon scattering equations}

\subsection{$\protect\bbox{\gamma^3}$He $\protect\bbox{\rightarrow N\!N\!N}$}

Photodisintegration of the three-nucleon bound state into three free nucleons is
described by the electromagnetic $^3$He $\rightarrow N\!N\!N$ transition current
$j_0^\mu$ consisting of all possible diagrams for this process
(``photodisintegration'' here means disintegration due to either an
on-mass-shell or an off-mass-shell photon, so the case of electrodisintegration
is included). By comparison, the gauged vertex function $\Phi_1^\mu$ consists of
all possible diagrams for photodisintegration where nucleons 2 and 3 are last to
interact and where no photons are attached to the outgoing nucleons. As we
already have an equation for $\Phi_1^\mu$, \eq{Phi_1^mu_X}, all that is
necessary to obtain $j_0^\mu$ is to add the missing terms in
$\Phi_1^\mu$. Indeed, we can immediately write down that
\be
j_0^\mu=P_c\left( \Phi_1^\mu + \sum_{i=1}^3\Gamma_i^\mu d_i \Phi_1\right) 
\eqn{j_0^mu}
\ee
where $P_c$ is the operator which sums over all the cyclic permutations of the
three particle labels. The role of $P_c$ is to include diagrams where nucleons
other than 2 and 3 are last to interact.  The term $P_c \sum_i\Gamma_i^\mu
d_i\Phi_1$ consists of all possible diagrams where photons are attached to the
final state external legs.

Denoting the three-particle Feynman propagator by $G_0$,
\be
G_0 = d_1 d_2 d_3,
\ee 
then
\be
G_0^{-1}G_0^\mu 
= G_0^{-1}\left(d_1^\mu d_2d_3+d_1d_2^\mu d_3+d_1d_2d_3^\mu\right)
=\sum_{i=1}^{3} \Gamma_i^\mu d_i,
\ee
and \eq{j_0^mu} can also
be written as
\be
j_0^\mu= P_c G_0^{-1}\left[ G_0\Phi_1\right]^\mu
= P_c \left( G_0^{-1}G_0^\mu\Phi_1+\Phi_1^\mu\right),  \eqn{j_0^mu-gauged}
\ee
indicating that $j_0^\mu$ can be obtained directly by gauging the quantity
$G_0\Phi_1$. This is just what one might expect since $G_0\Phi_1$ corresponds to
all possible diagrams for the $^3$He$\rightarrow N\!N\!N$ process where nucleons
(23) are last to interact. In this respect, it is interesting to note that
although $^3$He$\rightarrow N\!N\!N$ is not a possible physical process, it can
nevertheless be gauged to yield a physical electromagnetic process. It is also
worth pointing out that although we gauge on-mass-shell propagators when they
correspond to internal lines, only Feynman propagators are used in the
gauging of the external lines in \eq{j_0^mu-gauged}. This is not inconsistent
with the spectator approach, it preserves gauge invariance \cite{G3d}, and
it avoids introduction of on-mass-shell $\delta$-function like singularities
into the physical photodisintegration amplitude.

\subsection{$\protect\bbox{Nd\rightarrow \gamma Nd}$}

We can obtain the amplitude for the process $Nd\rightarrow \gamma Nd$ by gauging
the scattering amplitude for $Nd\rightarrow Nd$. Thus our first task is to
derive an expression for this amplitude.

From \eq{Phi_1^mu_X} it is clear that the quantity
$X=(1+t_1\delta_2d_3P_{12})^{-1}$ describes all possible perturbation graphs for
the process \NNNNNN\ where the 13 pair is the first and the 23 pair is the last
to interact (see also \fig{xfig}). By taking appropriate residues of $X$ we can
therefore obtain any scattering amplitude involving three nucleons, including
the one for $Nd$ elastic scattering. This we now proceed to do.

As seen explicitly in \fig{xfig}, the second iteration of either of the \eqs{X}
yields a connected graph. We can thus write
\be
X=(1+t_1\delta_2d_3P_{12})^{-1}=1-t_1\delta_2d_3P_{12} + X_c
\ee
where
\be
X_c =  (1+t_1\delta_2d_3P_{12})^{-1}t_1\delta_2d_3P_{12}t_1\delta_2 d_3P_{12}
\ee
is the connected part of $X$. Using the fact that
\be
t_1\delta_2d_3P_{12}t_1\delta_2d_3P_{12}=t_1d_3P_{12}
t_1P_{12}\delta_1\delta_2d_3 
\ee
and
\be
(1+t_1\delta_2d_3P_{12})^{-1}t_1=t_1(1+\delta_2d_3P_{12}t_1)^{-1},
\ee
we may write
\be
X=(1+t_1\delta_2d_3P_{12})^{-1}=1-t_1\delta_2d_3P_{12}+T_cP_{12}\delta_1
\delta_2d_3  \eqn{XTc}
\ee
where
\be
T_c=t_1(1+\delta_2d_3P_{12}t_1)^{-1}d_3P_{12}t_1     \eqn{T_c}
\ee
is the connected part of the scattering amplitude for \NNNNNN\ where the 23 pair
is both first and last to interact. It is easy to see that the corresponding
Bethe-Salpeter amplitude is given by
\be
T^{\mbox{\scriptsize BS}}_c=t_1(1+d_2d_3P_{12}t_1)^{-1}d_3P_{12}t_1   
\eqn{T_c^BS}
\ee
showing explicitly that the spectator equation expression of \eq{T_c} can be
obtained from the Bethe-Salpeter expression of \eq{T_c^BS} by replacing the
spectator particle's propagators by the on-mass-shell propagator in each
term of the perturbation series for $T^{\mbox{\scriptsize BS}}_c$.

The two-nucleon t-matrix $t_1$ contains the deuteron bound state pole. In the
vicinity of this pole we have that
\be
t_1(P;p_2,k_2)\sim i \frac{\phi_{23}(p_2)\bar{\phi}_{23}(k_2)}
{P^2-M_d^2}=i\frac{\phi_{23}\bar{\phi}_{23}}{P^2-M_d^2}  \eqn{33'}
\ee
where $P$ is the deuteron four-momentum, $M_d$ is the deuteron mass, and
$\phi_{23}$ is the deuteron vertex function for nucleons 2 and 3.
The scattering amplitude $T_{dd}$ for $Nd\rightarrow Nd$ is then obtained from
\eq{T_c} by taking left and right residues at the deuteron pole:
\be
T_{dd}=\bar{\phi}_{23}(1+\delta_2d_3P_{12}t_1)^{-1}d_3P_{12}\phi_{23} .
\eqn{T_dd}
\ee

The electromagnetic $Nd\rightarrow Nd$ transition current $j^\mu_{dd}$ that
describes the process $Nd\rightarrow \gamma Nd$ can now be obtained as in the
four-dimensional case \cite{G4d} by gauging $d_1T_{dd}d_1$. Defining
\be
Y = (1+\delta_2d_3P_{12}t_1)^{-1},       \eqn{Y_def}
\ee
we therefore have that
\bea
j^\mu_{dd}&=&d_1^{-1}\left(\bar{\phi}_{23}d_1Y
d_3P_{12}d_1\phi_{23}\right)^\mu d_1^{-1}\nn[3mm]
&=&\left(\bar{\phi}_{23}Yd_3P_{12}\phi_{23}\right)^\mu+
\bar{\phi}_{23}\Gamma^\mu_1d_1Yd_3P_{12}\phi_{23}+
\bar{\phi}_{23}Yd_3d_2\Gamma^\mu_2P_{12}\phi_{23}   \eqn{j^mu_dd}
\eea
where the first term on the r.h.s. is given by
\be
\left(\bar{\phi}_{23}Yd_3P_{12}\phi_{23}\right)^\mu
=\bar{\phi}_{23}^\mu Yd_3P_{12}\phi_{23}
+\bar{\phi}_{23}Y^\mu d_3P_{12}\phi_{23}
+\bar{\phi}_{23}Yd_3^\mu P_{12}\phi_{23}
+\bar{\phi}_{23}Yd_3P_{12}\phi_{23}^\mu.
\ee
The gauged vertex functions $\phi_{23}^\mu$ and $\bar{\phi}_{23}^\mu$ can be
obtained by gauging the two-body bound state equations
\be
\phi_{23}=\frac{1}{2}v_1\delta_2d_3\phi_{23},\hspace{2cm}
\bar{\phi}_{23}=\frac{1}{2}\bar{\phi}_{23}\delta_2d_3v_1 .
\ee
Using the equations for $t_1$, \eqs{t_1}, one easily obtains that
\be
\phi_{23}^\mu
=\frac{1}{2}\left(1+\frac{1}{2}t_1\delta_2d_3\right)
(v_1\delta_2d_3)^\mu\phi_{23},
\hspace{1cm}
\bar{\phi}_{23}^\mu=\frac{1}{2}\bar{\phi}_{23}(\delta_2d_3v_1)^\mu
\left(1+\frac{1}{2}\delta_2d_3t_1\right) .
\ee
To obtain an expression for $Y^\mu$, we first note that $Y$ satisfies the
equations
\be
Y=1-\delta_2d_3P_{12}t_1Y; \hspace{2cm} Y=1-Y\delta_2d_3P_{12}t_1. \eqn{Y}
\ee
Gauging either of these  equations then gives
\bea
Y^\mu &=& - Y\left(\delta_2d_3P_{12}t_1\right)^\mu Y\nn
&=&- Y\left(\delta_2^\mu d_3P_{12}t_1+\delta_2d_3^\mu P_{12}t_1
+\delta_2d_3P_{12}t_1^\mu \right) Y.
\eea
In this way the transition current $j_{dd}^\mu$ is completely determined in
terms of one- and two-body input quantities. Note that our expression for
$j_{dd}^\mu$ is in terms of the quantity $Y$ rather than the $X$ introduced
earlier.  Yet it turns out that once the integrals over the fourth components
are taken in the expression for $j_{dd}^\mu$, then it is seen that the use of
$X$ or $Y$ in \eq{j^mu_dd} is completely equivalent. This is discussed in the
Appendix where we show how our four-dimensional expressions of the spectator
approach are reduced to three-dimensional forms suitable for numerical
calculations.

\subsection{$\protect\bbox{\gamma^3}$He $\protect\bbox{\rightarrow Nd}$}

To find the $^3$He $\rightarrow Nd$ electromagnetic transition current
$j^\mu_d$, it would be natural to simply take the left residue of the $^3$He
$\rightarrow N\!N\!N$ electromagnetic transition current $j^\mu_0$, given in
\eq{j_0^mu}, at the two-body bound state pole of nucleons 2 and 3. Although this
is straightforward, one can obtain exactly the same result in an even simpler
way by gauging the scattering amplitude $T_d$ for the off-shell process $^3$He
$\rightarrow Nd$. The expression for $T_d$ is easily found from the the bound
state equation for $\Phi_1$ in \eq{Phi_1} by taking the left residue at the
two-body bound state pole:
\be
T_d = -\bphi_{23}\delta_2d_3P_{12}\Phi_1    \eqn{T_d}.
\ee
To make sure that one includes the case where photons are attached to the free
nucleon in the final $Nd$ state (particle 1), it is sufficient to gauge $d_1T_d$
and then multiply from the left by the inverse propagator $d_1^{-1}$ at the end.
Thus the electromagnetic transition current $j_d^\mu$ which describes the
physical process $\gamma^3$He $\rightarrow Nd$ is given by

\bea
j_d^\mu &=& -d_1^{-1}(d_1\bphi_{23}\delta_2d_3P_{12}\Phi_1)^\mu\nn
&=& 
-\left(\Gamma_1^\mu d_1\bphi_{23}\delta_2d_3
+\bphi_{23}^\mu\delta_2d_3 
+\bphi_{23}\delta_2^\mu d_3
+\bphi_{23}\delta_2d_3^\mu \right)P_{12}\Phi_1
-\bphi_{23}\delta_2d_3P_{12}\Phi_1^\mu,
\eea
where all quantities have been specified above.
\section{Summary}

We have derived relativistic three-dimensional integral equations describing the
interaction of the three-nucleon system with an external electromagnetic
field. In particular, we have presented expressions for the three-nucleon bound
state electromagnetic current, as well as for the transition currents describing
the scattering processes $\gamma ^3$He$\rightarrow N\!N\!N$, $Nd\rightarrow
\gamma Nd$, and $\gamma ^3$He$\rightarrow Nd$. Our equations are gauge invariant
and conserve charge. More importantly, gauge invariance and charge conservation
are achieved in the theoretically correct fashion; namely, by the attachment of
photons to all possible places within the strong interaction model of the 
three nucleons.

The achievement of these results was made possible by the recent development of
the gauging of equations method \cite{G4d}. Previously this method was used to
generate a four-dimensional gauge invariant description of the three-nucleon
system and its electromagnetic currents. Here we applied the same method to what
in principle is an even more challenging problem, namely, the gauging of the
spectator equations for the three-nucleon system \cite{Gross}. The extra
difficulty in this case comes from the question of how to choose the spectator
particles once the gauging of the four-dimensional equations is done. We solved
this problem by (i) working in terms of Faddeev components, and (ii) by
introducing the idea of an on-mass-shell nucleon propagator $\delta$ in order to
express the three-nucleon spectator equations in a four-dimensional form
\cite{G3d}. Once in this form, the spectator equations were then gauged
directly, in this way allowing the gauging method itself to determine the
spectator particles in the final gauged equations.

An important ingredient in our gauged equations is the gauged on-mass-shell
propagator $\delta^\mu$.  The question of how to construct a form for
$\delta^\mu$ that satisfies both the Ward-Takahashi identity and the Ward
identity was previously answered in Ref.~\cite{G3d} and \cite{delta^mu}.  As
both $\delta$ and $\delta^\mu$ contain on-mass-shell $\delta$-functions, our
gauged four-dimensional equations can be reduced to a three-dimensional
form. The details of this reduction were presented in the Appendix. As a result,
we have brought together all the theoretical results that are necessary for a
practical calculation of the electromagnetic processes of the three-nucleon
system.

\acknowledgments

The authors would like to thank the Australian Research Council for their
financial support.

\appendix
\section*{}

In the main part of this paper, all our results have been expressed in terms of
four-dimensional integrals despite the presence of $\delta$-functions which
could allow us to reduce the integrals to three-dimensional ones. This has been
done specifically so that we can follow the gauging procedure introduced
in Ref.\cite{G4d} for four-dimensional integral equations. Our final results,
however, are three-dimensional, and it is the purpose of this appendix to write
out some of the obtained expressions in a purely three-dimensional form.

We begin with the bound state equation of \eq{Phi_1}. This symbolic
equation represents the four-dimensional integral equation
\be
\Phi_1^Q(p_1,p_2) = - \int \frac{d^4k_2}{(2\pi)^4} t(Q-p_1;p_2,k_2)
\delta(k_2) d(Q-p_1-k_2) P_{12}\Phi_1^Q(p_1,k_2).  \eqn{Phi_1_ex}
\ee
Because of the presence of the on-mass-shell propagator $\delta(k_2)$, the
integral over $dk_2^0$ may be done trivially.
Setting the momenta $p_1$ and $p_2$ to be on mass shell we then obtain
the three-dimensional equation
\be
\Phi_1^Q(\bp_1,\bp_2) = - \int \frac{d^3k_2}{(2\pi)^3 }
t(Q-\bp_1;\bp_2,\bk_2)\frac{\Lambda(\bk_2)}{2\omega_{k_2}} d(Q-\bp_1-\bk_2)
P_{12}\Phi_1^Q(\bp_1,\bk_2)                          \eqn{Phi_1_on}
\ee
where $\omega_{k}=\sqrt{{\bfk}^2+m^2}$ and $\bar{k} =
\left(\omega_{k},\bfk\right)$. Although this equation is three-dimensional, the
quantities involved still retain their Dirac spinor structure. Thus, for
example, $\Phi^Q_1(\bp_1,\bp_2)$ consists of a direct product of three Dirac
spinors, one for each nucleon, while $t(Q-\bp_1;\bp_2,\bk_2)$ is a $16\times 16$
matrix. For the on-mass-shell nucleons we may eliminate the Dirac spinor
structure by appropriate multiplication by the Dirac spinors $u$ or $\bu$. We
therefore define
\be
\tPhi_1^Q(\bfp_1\alpha_1,\bfp_2\alpha_2) =
\bu(\bfp_1,\alpha_1)\bu(\bfp_2,\alpha_2)\Phi_1^Q(\bp_1,\bp_2) 
\ee
and
\be
\ttt(Q-\bp_1;\bfp_2\alpha_2;\bfk_2\beta_2)=
\bu(\bfp_2,\alpha_2)t(Q-\bp_1;\bp_2,\bk_2)u(\bfk_2,\beta_2).
\ee
Since
\be
\Lambda(\bp) = \pslash + m = 2m\sum_\alpha u(\bfp,\alpha)\bu(\bfp,\alpha),
\ee
where the normalization of the Dirac spinors is given by
$\bu(\bfp,\alpha)u(\bfp,\beta)=\delta_{\alpha\beta}$, \eq{Phi_1_on} can be
transformed to the equation
\be
\tPhi_1^Q(\bfp_1\alpha_1,\bfp_2\alpha_2) = - \sum_{\beta_2}\int
\frac{d^3k_2}{(2\pi)^3}\frac{m}{\omega_{k_2}}
\ttt(Q-\bp_1;\bfp_2\alpha_2,\bfk_2\beta_2)
\td(Q,\bfp_1,\bfk_2) P_{12}\tPhi_1^Q(\bfp_1\alpha_1,\bfk_2\beta_2)
                        \eqn{Phi_1_3dx}
\ee
where
\be
\td(Q,\bfp_1,\bfk_2)=d(Q-\bp_1-\bk_2),
\ee
with the integral being taken over the Lorentz invariant phase space volume
$\displaystyle \frac{d^3k_2}{(2\pi)^3}\frac{m}{\omega_{k_2}}$.  We shall write
this three-dimensional equation in symbolic form as
\be
\tPhi_1^Q = - \ttt_1\td_3 P_{12}\tPhi_1^Q .   \eqn{Phi_1_3d}
\ee
Note that \eq{Phi_1_3d} can be considered as an operator expression in
three-dimensional momentum space and is written in terms of tilde quantities to
distinguish it from \eq{Phi_1} which is a symbolic equation in four-dimensional
momentum space.

Similarly, the first scattering equation of \eqs{X} is a symbolic
expression for a four-dimensional integral equation which, after the trivial
integration over $dk_2^0$, gives
\bea
X(Q;p_1p_2;q_1q_2) &=& (2\pi)^8\delta^4(p_1-q_1)\delta^4(p_2-q_2)\nn
 &-& \int \frac{d^3k_2}{(2\pi)^3 } t(Q-p_1;p_2,\bk_2)
\frac{\Lambda(\bk_2)}{2\omega_{k_2}}  d(Q-p_1-\bar{k}_2)
P_{12}X(Q;p_1\bk_2,q_1q_2).
\eea
Putting $p_1$ and $p_2$ on mass shell in this equation, it can be noticed that
the inhomogeneous term becomes
\be
(2\pi)^8\delta^4(\bp_1-q_1)\delta^4(\bp_2-q_2)
=(2\pi)^8\delta^3(\bfp_1-\bfq_2)\delta^3(\bfp_2-\bfq_2)
\delta(\bq_1^0-q_1^0)\delta(\bq_2^0-q_2^0).
\ee
This in turn implies that
\be
X(Q;\bp_1\bp_2;q_1q_2) = X'(Q;\bp_1\bp_2;\bq_1\bq_2)\,
\frac{m}{\omega_{q_1}}\frac{m}{\omega_{q_2}}(2\pi)^2
\delta(\bq_1^0-q_1^0)\delta(\bq_2^0-q_2^0)   \eqn{X2on}
\ee
where $X'(Q;\bp_1\bp_2;\bq_1\bq_2)$ satisfies the three-dimensional equation
\bea
X'(Q;\bp_1\bp_2;\bq_1\bq_2)
&=& (2\pi)^6\, \frac{\omega_{q_1}}{m}\frac{\omega_{q_2}}{m}
\delta^3(\bfp_1-\bfq_1)\delta^3(\bfp_2-\bfq_2)\nn
 &-& \int \frac{d^3k_2}{(2\pi)^3 } t(Q-\bp_1;\bp_2,\bk_2)
\frac{\Lambda(\bk_2)}{2\omega_{k_2}} d(Q-\bp_1-\bk_2)
P_{12}X'(Q;\bp_1\bk_2,\bq_1\bq_2)  .   \eqn{Xon}
\eea
Note that \eq{X2on} provides the key result that has enabled us to eliminate the
zero-component $\delta$-functions from the inhomogeneous term.

We now eliminate the Dirac spinor structure of on-mass-shell particles by
defining
\be
\tX(Q;\bfp_1\alpha_1,\bfp_2\alpha_2;\bfq_1\beta_1,\bfq_2\beta_2)=
\bu(\bfp_1,\alpha_1)\bu(\bfp_2,\alpha_2)X'(Q;\bp_1\bp_2,\bq_1\bq_2)
u(\bfq_1,\beta_1)u(\bfq_2,\beta_2)
\ee
In this way \eq{Xon} gets transformed into a three-dimensional equation,
similar to \eq{Phi_1_3dx}:
\bea
\lefteqn{\tX(Q;\bfp_1\alpha_1,\bfp_2\alpha_2;\bfq_1\beta_1,\bfq_2\beta_2)=
\delta_{\alpha_1\beta_1}\delta_{\alpha_2\beta_2}
(2\pi)^6\, \frac{\omega_{q_1}}{m}\frac{\omega_{q_2}}{m}
\delta^3(\bfp_1-\bfq_1)\delta^3(\bfp_2-\bfq_2)}\nn
&-& \sum_{\gamma_2}\int \frac{d^3k_2}{(2\pi)^3}\frac{m}{\omega_{k_2}}
\ttt(Q-\bp_1;\bfp_2\alpha_2,\bfk_2\gamma_2) \td(Q,\bfp_1,\bfk_2)
P_{12}\tX(Q;\bfp_1\alpha_1,\bfk_2\gamma_2;\bfq_1\beta_1,\bfq_2\beta_2).
\eqn{tX}
\eea
Since the momentum phase space integration volume includes a factor
$(2\pi)^{-3}m/\omega$, the inhomogeneous term in this equation acts like a unit
operator in momentum space. Thus we can write this equation in symbolic form as

\be
\tX = 1 - \ttt_1\td_3P_{12}\tX.  \eqn{X=1-tX_3d}
\ee
It then follows that $\tX=(1+\ttt_1\td_3P_{12})^{-1}$ and therefore
\be
\tX = 1 - \tX\ttt_1\td_3P_{12}.  \eqn{X=1-Xt_3d}
\ee
Alternatively, one can show \eq{X=1-Xt_3d} by starting from the numerical
form of the four-dimensional equation $X=1-Xt_1\delta_2d_3$, setting the
final momenta $p_1$ and $p_2$ to be on mass shell, and using \eq{X2on} as
before. This time, however, one also needs to use the relation
\be
\bu(\bfp_1,\alpha_1)\bu(\bfp_2,\alpha_2)X'(Q;\bp_1\bp_2,\bq_1\bq_2)
=\sum_{\beta_1\beta_2}
\tX(Q;\bfp_1\alpha_1,\bfp_2\alpha_2;\bfq_1\beta_1,\bfq_2\beta_2)
\bu(\bfq_1,\beta_1)\bu(\bfq_2,\beta_2)  \eqn{uuX'=tXuu}
\ee
which can be proved by using \eqs{Xon} and (\ref{tX}) to show that each side of
\eq{uuX'=tXuu} satisfies the same integral equation.

Note that the bound state and scattering equations of \eq{Phi_1_3d} and
\eq{X=1-tX_3d} have the same kernel and can therefore be calculated using
similar numerical codes.  It would therefore be convenient to express the
amplitude for $Nd$ scattering, $T_{dd}$, in terms of $\tX$ rather than leaving
it in terms of $Y$ in which it is given in \eq{T_dd}.  To this end we first
write $Y$ in terms of $X$ by using the definitions of \eq{X_def} and \eq{Y_def}:
\be
Y=1-\delta_2d_3P_{12}Xt_1  .   \eqn{Y=1-X}
\ee
Then the $Nd$ amplitude is given by
\bea
T_{dd}&=&\bar{\phi}_{23}Yd_3P_{12}\phi_{23} \eqn{T_ddY}\\
&=&\bar{\phi}_{23}(1-\delta_2d_3P_{12}Xt_1)d_3P_{12}\phi_{23}. \eqn{T_ddX}
\eea
As particle 1 in the final state is on mass shell, we see that both particles 1
and 2 to the left of $X$ in the latter equation are on mass shell. We can
therefore use \eq{X2on} and \eq{uuX'=tXuu} in \eq{T_ddX} to write the physical
$Nd$ scattering amplitude $\tT_{dd}=\bu_1T_{dd}u_1$ in the three-dimensional
form
\be
\tT_{dd}=\bar{\tphi}_{23}(1-\td_3P_{12}\tX\ttt_1)\td_3P_{12}\tphi_{23}
\ee
where $\tphi_{23}=\bu_2\phi_{23}$, and $\bar{\tphi}_{23}=\bphi_{23}u_2$.
Writing this equation as
\be
\tT_{dd}=\bar{\tphi}_{23}\td_3P_{12}(1-\tX\ttt_1\td_3P_{12})\tphi_{23},
\ee
we may then use \eq{X=1-Xt_3d} to obtain that
\be
\tT_{dd}=\bar{\tilde{\phi}}_{23}\tilde{d}_3P_{12}\tilde{X} 
\tilde{\phi}_{23}    .
\ee

We now show how the physical transition current
$\tj^\mu_{dd}=\bu_1j^\mu_{dd}u_1$ can be similarly expressed in terms of
three-dimensional expressions involving $\tilde{X}$. In terms of $Y$, the
expression for $j^\mu_{dd}$ was given in \eq{j^mu_dd} which we can write in the
form
\bea
j^\mu_{dd}&=&
\bar{\phi}_{23}Yd_3P_{12}\left(\Gamma_3^\mu d_3\phi_{23}+\phi_{23}^\mu\right)
+\bar{\phi}_{23}^\mu Yd_3P_{12}\phi_{23}
-\bar{\phi}_{23}Yd_3P_{12}(\Gamma_3^\mu d_3t_1+t_1^\mu)
\delta_1 Yd_3P_{12}\phi_{23}\nn
&-&\bar{\phi}_{23}Yd_3P_{12}t_1\delta_1^\mu Yd_3P_{12}\phi_{23}
+\bar{\phi}_{23}\Gamma^\mu_1d_1Yd_3P_{12}\phi_{23}
+\bar{\phi}_{23}Yd_3P_{12}d_1\Gamma^\mu_1\phi_{23} .   \eqn{nnn}
\eea
Every $Y$ in this equation comes in the combination $Yd_3P_{12}$, just as in
\eq{T_ddY} for $T_{dd}$. In each of the first three terms on the r.h.s. of
\eq{nnn} (upper line), every $Y$ has the first particle on its left and the
second particle on its right restricted to be on mass shell. Thus we may proceed
as for \eq{T_ddY} and replace our expressions by three-dimensional ones where
$Yd_3P_{12}$ is replaced by $\td_3P_{12}\tX$. For example the first term becomes
\be
\bu_1\bar{\phi}_{23}Yd_3P_{12}\left(\Gamma_3^\mu d_3\phi_{23}
+\phi_{23}^\mu\right)u_1=\bar{\tphi}_{23}\td_3P_{12}\tX
\left(\Gamma_3^\mu \td_3\tphi_{23}+\tphi_{23}^\mu\right).
\ee
On the other hand, in the last three terms of \eq{nnn} every $Y$ has either an
off mass shell first particle on its left or an off mass shell second particle
on its right.  For example, in the last term of Eq.~(\ref{nnn}) particle 2 to
the right of $Y$ is off mass shell. Proceeding nevertheless as before, we have
that
\bea
\bu_1\bar{\phi}_{23}Yd_3P_{12}d_1\Gamma^\mu_1\phi_{23}u_1
&=&\bu_1\bar{\phi}_{23}(1-\delta_2d_3P_{12}Xt_1)d_3P_{12}d_1\Gamma^\mu_1
\phi_{23}u_1\nn
&=&\bar{\phi}_{23}d_3P_{12}d_1\Gamma^\mu_1u_1\tphi_{23}
-\bar{\tphi}_{23}d_3P_{12}\tX t^L_1
d_3P_{12}d_1\Gamma^\mu_1\tphi_{23}\nn \eqn{last_term}
\eea
where again we have used \eqs{X2on} and (\ref{uuX'=tXuu}) to reduce the
integrals to three-dimensions.  In the last equation, $t_1$ has only its left
particle 2 on mass shell, hence the notation $t^L_1$. Thus, although
Eq.~(\ref{last_term}) is three-dimensional and in terms of $\tX$, it cannot be
simplified further with the help of \eq{X=1-Xt_3d}. What can be done, however,
is to use the second of the equations for $t_1$, \eqs{t_1}, in order to express
$t^L_1$ in terms of $\ttt_1$ and $v^L_1$; indeed, it's easily seen
that $t^L_1$ is given by the three-dimensional equation
\be
t^L_1=v^L_1+\frac{1}{2}\ttt_1 \td_3 v^L_1.
\ee
With the help of \eq{X=1-Xt_3d}, this enables one to re-express \eq{last_term}
in terms of $v^L_1$.

A similar three-dimensional reduction holds for the
second last term in Eq.~(\ref{nnn}). The first term in the last line of
Eq.~(\ref{nnn}) contains the gauged on-mass-shell propagator $\delta_1^\mu$. As
seen from \eq{delta^mu}, $\delta_1^\mu$ has one of its legs on mass shell while
the other is off mass shell. Thus one of the terms $Yd_3P_{12}$ in the fourth
term of \eq{nnn} can be simplified by using \eq{X=1-Xt_3d}, while the other one
cannot.


%
%

%
%
\end{document}